\documentclass[aps,twocolumn,showpacs,preprintnumbers,amsmath,amssymb]{revtex4}
\usepackage{epsfig} 

 \newcommand{\re}{{\rm Re}}
 
 \newcommand{\Tr}{{\rm Tr}}

\newcommand{\be}{\begin{equation}}
\newcommand{\ee}{\end{equation}}
\newcommand{\bea}{\begin{eqnarray}}
\newcommand{\eea}{\end{eqnarray}}
\begin{document}

\title{Spin-3/2 baryons from an anisotropic lattice QCD action}
\author{Leming Zhou}
\author{Frank X. Lee}
\affiliation{Center for Nuclear Studies, Physics Department, The George Washington University, Washington DC 20052}

\date{\today}

\begin{abstract}
The mass spectrum of baryons in the spin-3/2 sector is 
computed in quenched lattice QCD using a tadpole-improved anisotropic action.
Both isospin 1/2 and 3/2 (the traditional decuplet) are considered, 
as well as members that contain strange quarks.
States with positive and negative parities are 
isolated by parity projection, while states with spin-3/2 and spin-1/2 
are separated by spin projection. 
The extent to which spin projection is needed is examined. 
The issue of optimal interpolating field is also investigated.
The results are discussed in relation to previous calculations and experiment.
\end{abstract}


  \maketitle

\section{Introduction}
\label{sec:intro}
The mass spectrum of hadrons represents a fundamental manifestation of 
the long-distance dynamics of quarks and gluons as governed by QCD. 
Non-perturbative calculations through numerical simulations on a
space-time lattice provide a method to obtain this spectrum from  first principles. 
The computation of the hadron spectrum using lattice 
QCD started in the early 1980's~\cite{Hamber:1981zn,Weingarten:1982jy}. 
The modern era in lattice QCD calculation of the hadron spectrum started 
with the results of the GF11 group~\cite{Butler:em}. The benchmark 
calculation of the quenched light hadron spectrum using the standard Wilson
action has been performed by the CP-PACS collaboration~\cite{Yoshie:ts,Aoki:1999yr}. 
For state-of-the-art computations using dynamical configurations 
that involve the study of baryon mass spectrum, see, for example, 
\cite{sesam99,ukqcd99,cppacs02,hpqcd04}.
For reviews on baryon mass spectrum, see, for example, 
\cite{craig03,derek04}.

Most of the lattice computation of the light hadron spectrum 
has been limited to the ground states.
It is important to extend the successes beyond the ground state. 
The rich structure of the excited baryon spectrum, as tabulated by the 
Particle Data Group~\cite{pdg04}, provides a fertile ground for exploring 
how the internal degrees of freedom in the nucleon are excited 
and how QCD works in a wider context. 
One example is the parity splitting in the low-lying $N^*$ spectrum.
The nucleon $N(938)$ has positive parity, while its negative parity excitation, $S_{11}(1535)$, 
has a much higher mass. 
The spontaneous chiral symmetry breaking in QCD is thought to be responsible for the splitting.
Without it, QCD would predict exact parity doubling in the baryon spectrum.
The study of the excited mass spectrum is a critical part of 
the experimental program at Jefferson Lab. 
Lattice QCD has a number of advantages in helping understand the $N^*$ spectrum.
One can systematically study the spectrum sector by sector, with the ability 
to dial the quark masses, to separate the parities exactly, to project out the spin components, 
and eventually to dissect the degrees of freedom in the QCD vacuum most responsible for the spectrum.
There have been a number of lattice calculations of the $N^*$ 
in the spin-1/2 sector~\cite{Lee:1998cx,Lee:2000hh, Sasaki:1999yh,Blum:2000cf,Sasaki:2001nf,
Sasaki:2001tg, Richards:2000hp,Gockeler:2001db,Melnitchouk:2002eg,bgr03,bgr06}. 

In this work, we focus on the spin-3/2 sector. In addition to the usual baryon decuplet with 
isospin-3/2 and spin-3/2, we study the isospin-1/2 and spin-3/2 family which
has only received limited attention so far. A preliminary study was reported in Ref.~\cite{Lee:2001ts}.
A calculation using the FLIC fermion by the Adelaide group was done in Ref.~\cite{Zanotti:2003fx}.
Other mehtods for constructing higher spin states have been proposed by the LHPC 
collaboration~\cite{Basak:2004hr,Basak:2005aq,Basak:2005ir}.
Here, we use a different interpolating field as the one used in Ref.~\cite{Zanotti:2003fx}.
We also extend the calculation to include states that contain the strange quark.
The goal is to establish the basic features in terms of spin-parity on the lattice.


\section{Calculation Details}
\label{sec:details}

Excited states composed of light constituents are both large in size and mass.
Their calculation imposes severe signal-to-noise problems. 
The use of an anisotropic lattice can help alleviate the problem.
A fine lattice in the temporal direction enables the 
correlator to be observed over many time slices at short separations, 
while the coarse spatial spacing allows large spatial volumes
the states demand. 
We use the anisotropic gauge action given in Ref.~\cite{Morningstar:1997ff}: 
\begin{equation}
\label{eq:gauge}
S_G = \beta \{ 
\frac{5}{3} \frac{\Omega_{\rm sp}}{\xi u_s^4}
+ \frac{4}{3} \frac{\xi \Omega_{\rm tp}}{u_s^2 u_t^2}
- \frac{1}{12} \frac{\Omega_{\rm sr}}{\xi u_s^6}
- \frac{1}{12}\frac{\xi \Omega_{\rm str}}{u_s^4 u_t^2}
\} ,
\end{equation}
where 
$
u_s= \langle \frac{1}{3} \re \Tr P_{s s^\prime} \rangle^{1/4}
$
is the spatial tadpole factor, $P_{s s^\prime} $ denoting the spatial plaquette.
$u_t$ is the temporal tadpole factor,  we set $u_t =$ 1 in this simulation.
$\Omega_C = \sum_C \frac{1}{3} \re \Tr (1-W_C)$, with $W_C$ denoting the 
path-ordered product of link variables along a closed contour $C$ on the lattice. 
$\Omega_{\rm sp}$ includes the sum over all spatial plaquettes on the lattice, 
$\Omega_{\rm tp}$ indicates the temporal plaquettes, $\Omega_{\rm sr}$ denotes the 
product of link variables about planar $2 \times 1$ spatial rectangular loops, 
and $\Omega_{\rm str}$ refers to the short temporal rectangles (one temporal link, 
two spatial). 

\begin{widetext}
For the quarks, the anisotropic D234 action of Ref.~~\cite{Alford:1996nx,Alford:1997yy} is 
used with the following Dirac operator,
\be
\label{eq:D234caction}
M_{D234} = m (1+\frac{1}{2}ram ) 
             + \sum_\mu \{ \gamma_\mu \Delta_\mu^{(1)} 
          - \frac{a^2}{6}  \gamma_\mu \Delta_\mu^{(3)}  
          + r  [ - \frac{a}{2}  \Delta_\mu^{(2)}
          - \frac{a}{4}  \sum_\nu \sigma_{\mu\nu} F_{\mu\nu}
	 + \frac{a^3}{24}  \Delta_\mu^{(4)} ] \}.
\ee
Here $\Delta_\mu^{(n)}$ is the $n^{th}$ order lattice covariant
derivative, 
$\Delta^{(3)}=\Delta^{(1)}\Delta^{(2)}=\Delta^{(2)}\Delta^{(1)}$,
and $\Delta^{(4)}=\Delta^{(2)}\Delta^{(2)}$.  
The terms proportional to $r$ are generated by a field redefinition
and thus represent a redundant operator.
With the help of the gauge-covariant first- and second-order
lattice derivatives,
\begin{equation}
\nabla_\mu \, \psi(x) = \frac{1}{2a_\mu} \left [
U_\mu(x) \, \psi(x+\mu) - U_\mu^\dagger(x-\mu) \, \psi(x-\mu) \right ]\, ,
\end{equation}
and
\begin{equation}
\Delta_\mu \, \psi(x) = \frac{1}{a_\mu^2} \left [
U_\mu(x) \, \psi(x+\mu) + U_\mu^\dagger(x-\mu) \, \psi(x-\mu) - 2 \,
\psi(x) \right ],
\end{equation}
the action can be cast into the standard form of
\begin{eqnarray}
S_q &=&
\sum_{x}\left\{{ \over }
\;\bar{\psi}(x)\psi(x)
\right. \nonumber \\ & &\qquad
- \kappa \sum_{\mu} \bar{\psi}(x)\,
\left[r\,{a_t^2\over a_\mu^2}+8c_\mu {a_t\over a_\mu}
-(1+2b_\mu)\,{a_t\over a_\mu}\,\gamma_\mu\right] \,
{U_\mu(x)\over u_{0,\mu}} \, \psi(x+\mu)
\nonumber \\ & &\qquad \left.
-\kappa \sum_{\mu} \bar{\psi}(x+\mu)\,{U^\dagger_\mu(x)\over u_{0,\mu}}
\left[r\,{a_t^2\over a_\mu^2}+8c_\mu{a_t\over a_\mu}
+(1+2b_\mu){a_t\over a_\mu}\,\gamma_\mu\right] \,
\psi(x)
\right. \nonumber \\ & &\qquad
+ \kappa \sum_{\mu} \bar{\psi}(x)\,
\left(2c_\mu{a_t\over a_\mu}-b_\mu {a_t\over a_\mu}\,\gamma_\mu\right) \,
{U_\mu(x)U_\mu(x+\mu)\over u_{0,\mu}^2} \, \psi(x+2\mu)
\nonumber \\ & &\qquad \left.
+ \kappa \sum_{\mu}
\bar{\psi}(x+2\mu)\,{U^\dagger_\mu(x+\mu)U^\dagger_\mu(x)\over u_{0,\mu}^2}
\left(2c_\mu {a_t\over a_\mu}+b_\mu{a_t\over a_\mu}\,\gamma_\mu\right) \,
\psi(x)
\right. \nonumber \\ & &\qquad  \left.
- \kappa \sum_{\mu>\nu}
{r\,a_t^2\over a_\mu a_\nu u_{0,\mu}^2u_{0,\nu}^2} \,
\bar{\psi}(x)\,i\sigma_{\mu\nu}{F}_{\mu\nu}(x) \psi(x)
\right\}
\end{eqnarray}
\end{widetext}
With the specific choice of the factors
$b_\mu=\frac{1}{6}$, $c_\mu=\frac{ra_t}{24a_\mu}$, and $r=2/3$,
and an improved version of the field-strength operator free of $O(a^2)$ errors,
$F_{\mu \nu}^{({\rm clover})}$, and a relative $O(a^2)$ correction, 
\be
F_{\mu \nu}(x) \equiv F_{\mu \nu}^{(\rm clover)} (x)
-a^2 \frac{1}{6} ( \Delta_\mu^{(2)} \Delta_\nu^{(2)})
F_{\mu \nu}^{({\rm clover})} (x),
\ee
the action has only $O(a_t^4,a_s^4)$ classical errors.
This action consists of three types of interaction terms:
nearest-neighbor, next-nearest-neighbor, and the clover term.
Both gauge action and quark action have tadpole-improved tree-level coefficients 
to reduce unwanted quantum fluctuations.
The hopping parameter $\kappa$ is related to the bare parameters by
\begin{equation}
{1\over 2\kappa}=m_0a_t+\sum_{\mu}\left(r\,{a_t^2\over a_\mu^2}+
6c_\mu {a_t\over a_\mu}\right).
\end{equation}
In this calculation, we use an 10$^3 \times$ 30 
anisotropic lattice with anisotropy $\xi=a_s/a_t =3$.
The spatial lattice spacing $a_s\approx 0.24$ fm determined from the Sommer scale $r_0$. 
The lattice 
coupling $\beta=2.4$. In all, 100 configurations are analyzed. On each configuration 
9 quark propagators are computed using a multi-mass solver, with quark masses ranging 
from approximately 780 to 90 MeV. The nine $\kappa$ values are:
$
\kappa_{1-9} = 0.30, 0.31, 0.32, 0.33, 0.34, 0.345, 0.35, 0.355, 0.36
$
They correspond to pion mass in the range of
2.11 to 0.68 GeV, and the mass ratio $\pi/\rho$ from 0.95 to 0.65.
The strange quark mass corresponds to the seventh kappa value ($\kappa = 0.350$).  
The critical kappa value determined from $m_\pi^2$ is $\kappa_c=0.3705(3)$.
The source is located at $(x,y,z,t)=(1,1,1,2)$.
We use Dirichlet boundary conditions in the time direction.

A Gaussian-shaped, gauge-invariant smearing function~\cite{Gus90}
in spatial directions,
\begin{equation}
\left[1+\alpha \sum_{i=1,2,3}(U_{i}(x-i)+U_i^\dagger(x))\right]^N
\label{smear}
\end{equation}
was applied both at the source and at the sink
to increase the overlap with the states in question.
So for a given interpolating field operator, one can construct four types of
correlation functions with the source-sink combinations of local-local (LL),
smear-local (SL), local-smear (LS), and smear-smear (SS).
In Eq.(\ref{smear}), $\alpha$ is the coupling strength
at which the neighboring links are brought in, and $N$
is the number of iteration times.
We used $\alpha=0.25$ and $N=10$ in all cases. We found that the SL 
gives the best signal so the results presented in this work are from this combination.


We consider the full interpolating field with the quantum numbers 
$I(J^P)={1\over 2}\left({3\over 2}^+\right)$ as proposed in Ref.~\cite{Chung:cc},
\begin{equation}
\label{eq:chifull}
\chi^\mu = \epsilon_{abc}\left( {u^a}^{T}C\gamma_5\gamma_\rho d^b\right) 
\left(g^{\mu\rho}-{1\over 4}\gamma^\mu\gamma^\rho\right) \gamma_5 u^c.
\end{equation}
It satisfies the condition $\gamma_\mu \chi^\mu=0$ for spin-3/2 fields.
The superscript $T$ denotes transpose. The $C = \gamma_4 \gamma_2$ is 
the charge conjugation matrix.  The Dirac $\gamma $ matrices are Hermitian 
and satisfy $\{ \gamma_\nu, \gamma_\tau \} = 2 \delta_{\nu \tau}$, 
with $\sigma_{\nu \tau} = \frac{1}{2i} [\gamma_\nu, \gamma_\tau]$.
We follow the gamma-matrix notation of Sakurai~\cite{Sakurai}. 
The $\mu, \rho$ are Lorentz indices and summation over $\rho$ is implied. 
The antisymmetric $\epsilon_{abc}$ ensures that the state is color-singlet.
The interpolating fields of the spin-3/2 $\Sigma^*$ and $\Xi^*$ are
obtained by properly changing the quark field operators. For example, 
one can get the spin-3/2 $\Sigma^*$ state
interpolating field by substituting $d$ with $s$; and $\Xi^*$ by replacing $u$ with $s$.
This interpolating field has 5 terms as compared to the standard interpolating field for the 
nucleon,
\be
\chi(x) = \epsilon_{abc}\left( {u^a}^{T}(x) C\gamma_5 d^b(x) \right)  u^c(x).
\label{nuc1}
\ee
So computationally, it is 25 times more expensive. Furthermore, since the 
full 4x4 matrix in Dirac space (as opposed to only diagonal elements) 
is needed to carry out the spin projection described below,
an extra factor of 4 is needed, making this interpolating field 100 times more expensive 
than a standard nucleon mass calculation. 

Despite having an explicit parity by construction, the interpolating
field couples to both positive and negative parity states.
A parity projection is needed to separate the two. In the large Euclidean 
time limit, the two-point correlation function with Dirichlet boundary 
condition in the time direction and  at zero spatial momentum becomes
\bea
&&G_{\mu\nu}(t)  =  \sum_{\bf x} <0|\chi_\mu(x)\,\overline\chi_\nu(0)|0>  
 \nonumber  \\
&=& f_{\mu\nu}\left[\lambda_+^2 {\gamma_4 +1 \over 2} e^{- M_+ \, t}
+ \lambda_-^2 {-\gamma_4 +1 \over 2} e^{- M_- \, t} \right] 
\eea
where $f_{\mu\nu}$ is a function common to both terms.
The relative sign in front of $\gamma_4$ provides the solution: by taking the
trace of $G_{\mu\nu}(t)$ with $(1\pm\gamma_4)/4$, one can isolate 
$M_+$ and $M_-$, respectively. 

The interpolating field in Eq.~(\ref{eq:chifull})
couples to both spin-3/2 and spin-1/2 states. To project a pure 
spin-3/2 state from the correlation function $G_{\mu \nu}$, we 
use a spin-3/2 projection operator~\cite{Zanotti:2003fx,Benmerrouche:uc},
\begin{equation}
\label{eq:32proj}
P_{\mu \nu }({3/2}) = g_{\mu \nu } - \frac{1}{3} \gamma_\mu
\gamma_\nu  - \frac{1}{3 p^2} ( \gamma \cdot p \gamma_\mu p_\nu +
p_\mu \gamma_\nu \gamma \cdot p )
\end{equation}
The corresponding spin-1/2 state can be projected by applying the 
projection operator
\be
\label{eq:12proj}
P_{\mu \nu}({1/2}) = g_{\mu \nu} -P_{\mu \nu}({3/2})
\ee
The projection is done after the correlation functions are generated, 
with no need to generate new quark propagators at the source.
Only zero spatial momentum ($\vec{p}=0$) is considered in the projector.
To use this operator and retain all Lorentz components, one must 
calculate the full $4 \times 4$ matrix in Dirac and Lorentz space
of $G_{\mu \nu}(t)$.  
Using the projection, we have
\be
\label{eq:g331}
G_{\mu\nu}^{1/2} (t) = \sum \limits_{\lambda=1}^{4} G_{\mu \lambda} (t) 
P^{\lambda \nu}({1/2}) ,
\ee
\be
\label{eq:g332}
G_{\mu\nu}^{3/2} (t) = \sum \limits_{\lambda=1}^{4} G_{\mu \lambda} (t) 
P^{\lambda \nu}({3/2}) ,
\ee
They satisfy the relation
\begin{equation}
G_{\mu\nu}(t) = G_{\mu\nu}^{1/2}(t)+G_{\mu\nu}^{3/2}(t).
\label{sum}
\end{equation}
However, to extract the mass, only one diagonal pair of Lorentz indices is needed, 
reducing the amount of calculations required by a factor of 4. 
We calculate $G_{33}^{1/2}$ and $G_{33}^{3/2}$.


\section{Isospin-1/2 and Spin-3/2 Baryons}
\label{sec:nstar}

\begin{figure}
\parbox{.5\textwidth}{%
\centerline{\psfig{file=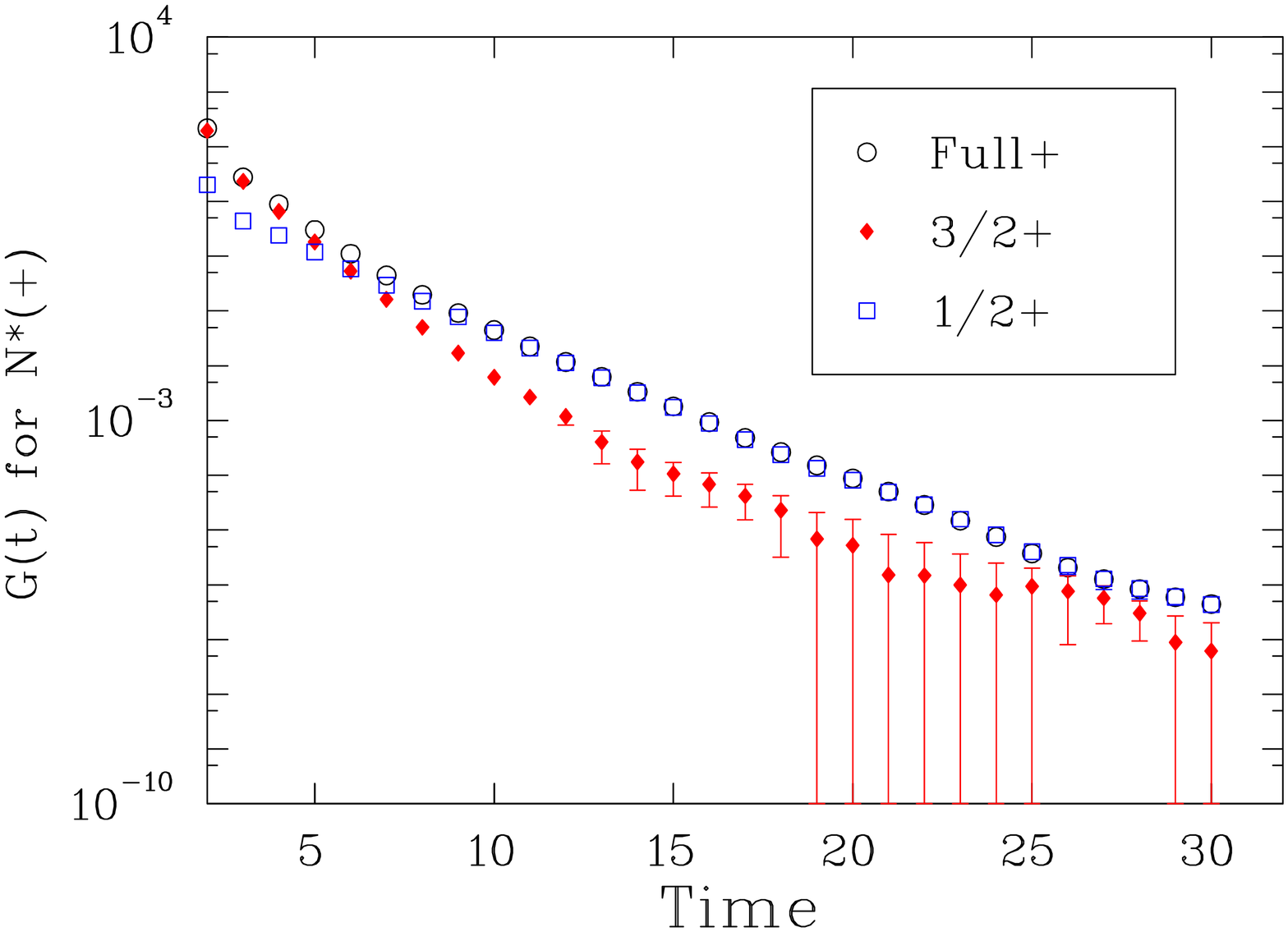,width=0.90\hsize,angle=90}}
}\hfill
\parbox{.5\textwidth}{%
\centerline{\psfig{file=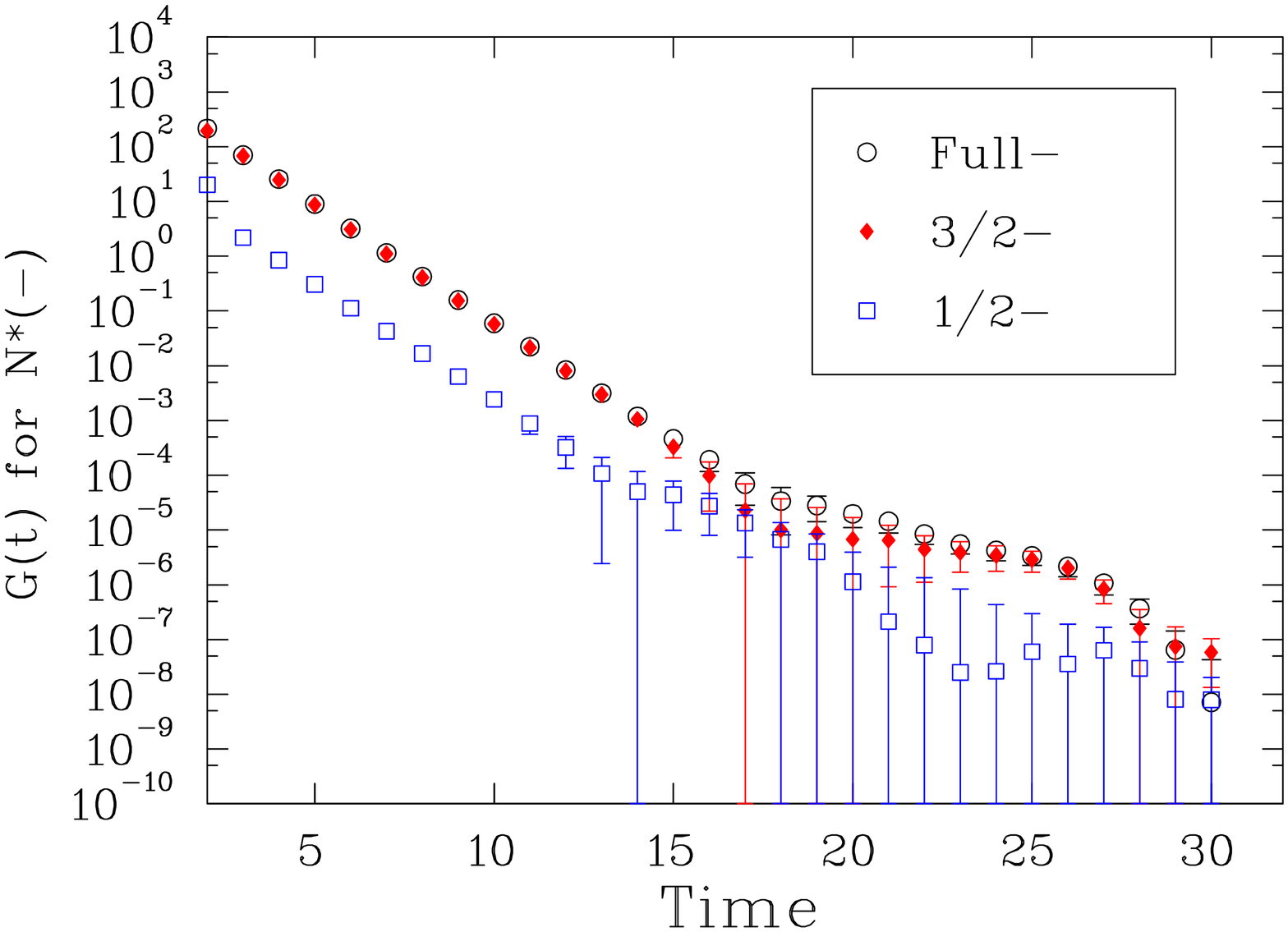,width=0.90\hsize,angle=90}}
}
\caption{The various correlation functions 
(unprojected, spin-3/2 projected, spin-1/2 projected) 
for the nucleon states in the positive-parity (top) and negative-parity (bottom) channels 
at the smallest quark mass ($\kappa=0.36$).}
\label{corr_n3tr}
\end{figure}

Fig.~\ref{corr_n3tr} demonstrates results for the
correlation function for both parities in the nucleon channel at the
smallest quark mass considered.  
In the positive-parity channel, 
one can see that this correlation function 
is almost completely dominated by the 1/2+ component. 
Spin projection reveals two different
exponentials from the spin-3/2 and spin-1/2 parts, with 
the spin-3/2 state being heavier than the spin-1/2 one (a steeper fall-off), 
in agreement with the ordering in experiment.  
The expected relation in Eq.~(\ref{sum}) is indeed satisfied numerically, 
providing a non-trivial check of the calculation.

A further check of the calculation is provided by the fact that 
the mass extracted from $G_{\mu \nu}^{1/2}(t)$ is degenerate with 
that from the conventional $G(t)$ for the nucleon ground state 
using the standard interpolating field in Eq.~(\ref{nuc1}). 
One can see that spin projection is crucial in this channel.
Without it, one would get a false signal for spin-3/2 since it is 
dominated by the spin-1/2 component state.  The large
error bars is a sign of sensitive cancellations in the projection procedure.  

The situation in the negative-parity channel is opposite, as shown in 
Fig.~\ref{corr_n3tr}.
Here the signal is dominated by the $3/2-$ state, so one would get a 
spin-3/2 signal without spin projection.
The results also show a similar fall-off for the $1/2-$ state and the $3/2-$ state,
in accord with the experimental states of $N^*(1535){1\over 2}^-$ and 
$N^*(1520){3\over 2}^-$ which are close to each other. 
We checked that the condition in Eq.~(\ref{sum}) is also satisfied.
\begin{figure}
\parbox{.5\textwidth}{%
\centerline{\psfig{file=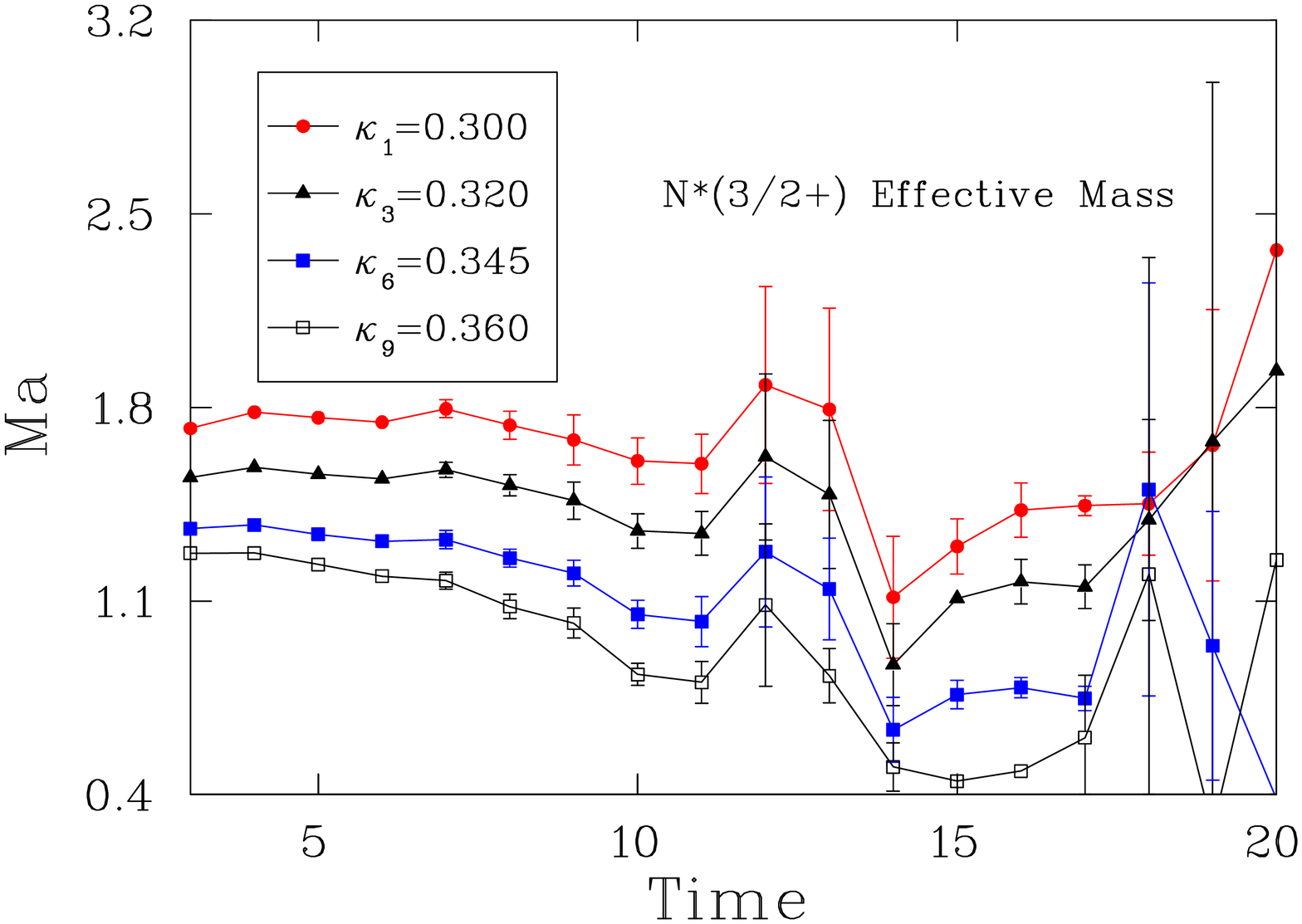,width=0.90\hsize,angle=90}}
}\hfill
\parbox{.5\textwidth}{%
\centerline{\psfig{file=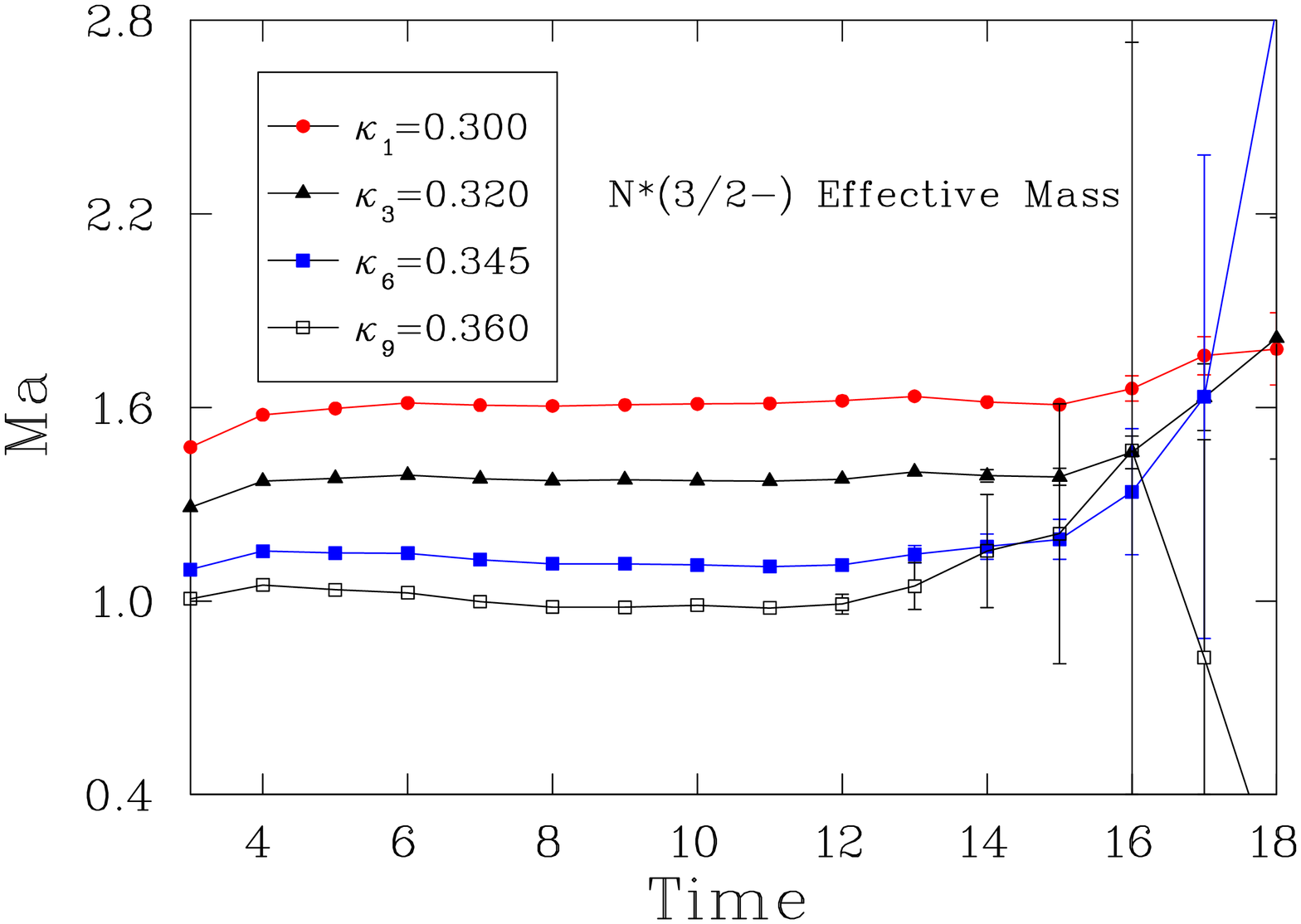,width=0.90\hsize,angle=90}}
}
\caption{Effective mass plot for the $N^*(3/2+)$ state (top) 
and $N^*(3/2-)$ state (bottom) at selected quark masses.}
\label{eMass_nuc32}
\end{figure}
\begin{figure}
\parbox{.5\textwidth}{%
\centerline{\psfig{file=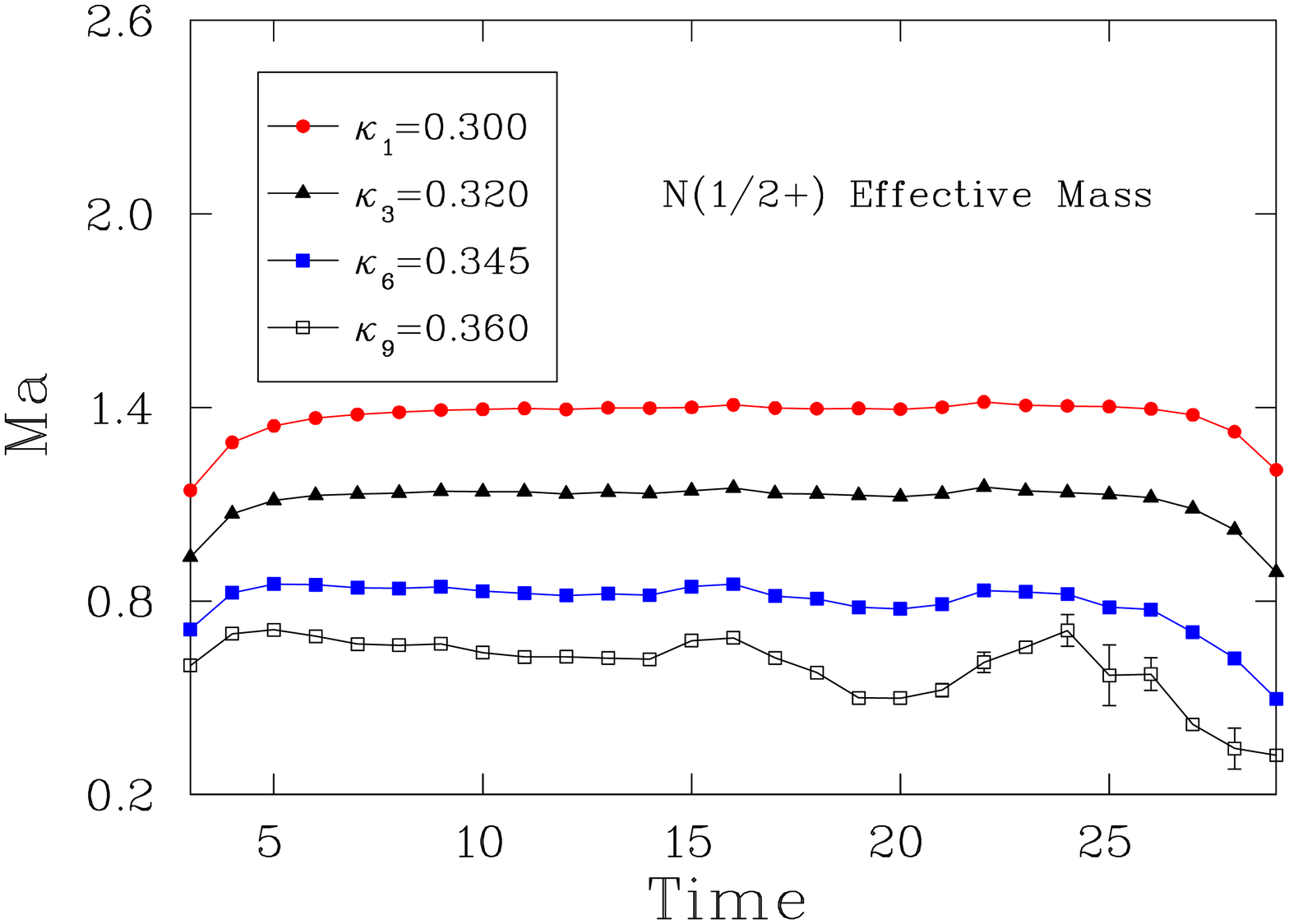,width=0.90\hsize,angle=90}}
}\hfill
\parbox{.5\textwidth}{%
\centerline{\psfig{file=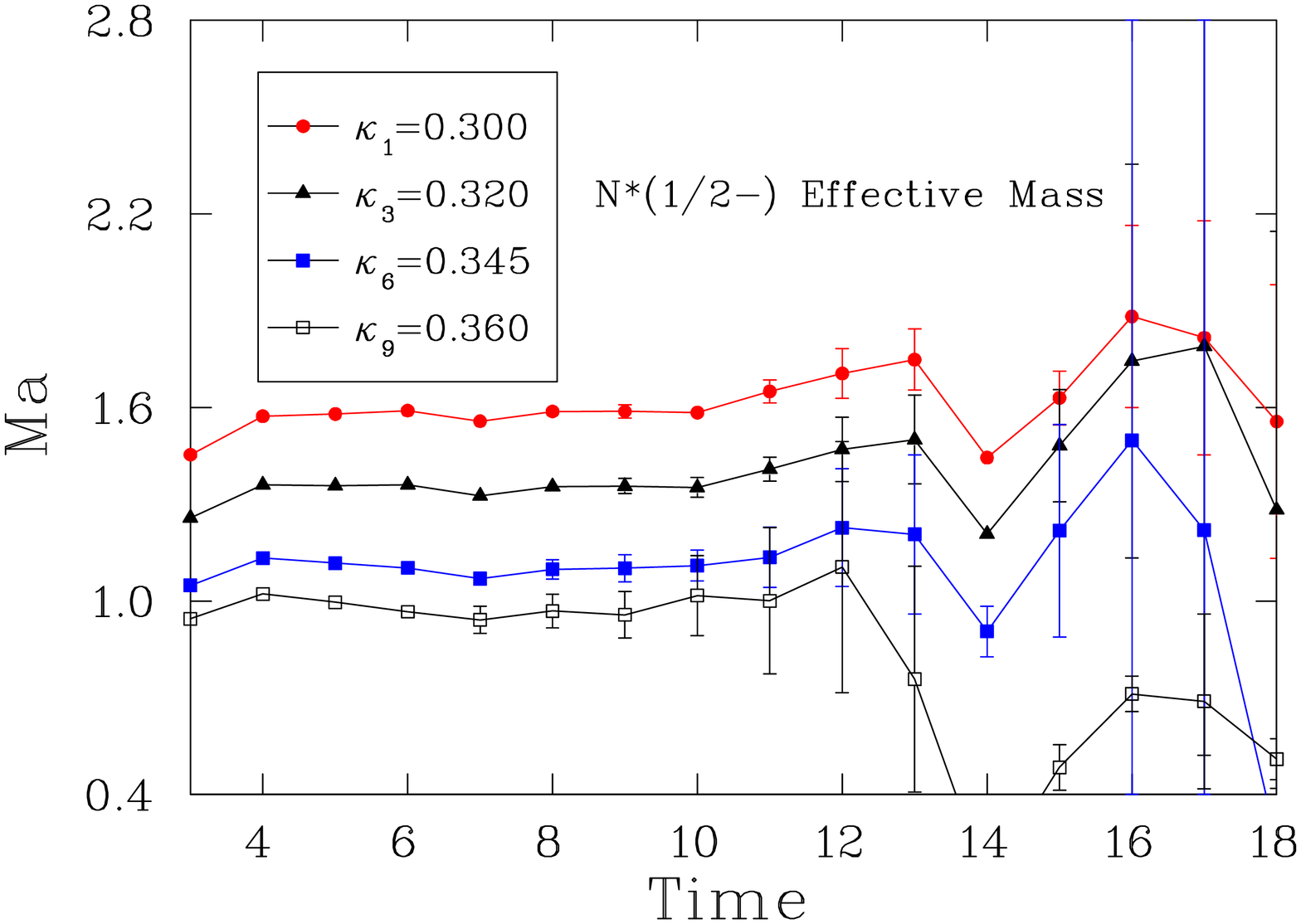,width=0.90\hsize,angle=90}}
}
\caption{Effective mass plot for the $N^*(1/2+)$ state (top) 
and $N^*(1/2-)$ state (bottom) at selected quark masses.}
\label{eMass_nuc12}
\end{figure}
According to Fig.~\ref{corr_n3tr}, 
we can get a rough idea about the range of time slices we should 
choose to extract the baryon masses. For example, to get good 
fitting result of $N^*(3/2+)$, we should choose time slices
earlier than 12, however, the mass of $N(1/2+)$ can be extracted from 
much later time slice. To find out more specific fit time window, we use
effective masses and extract the baryon masses from a plateau area.  
Fig.~\ref{eMass_nuc32} presents the effective masses in the spin-3/2 sector
at four quark masses that correspond to the heaviest quark mass, 
and two quark masses in the middle, and the lightest quark mass).
For $N^*(3/2+)$ the signal is weak because the correlation function 
for positive-parity is dominated by the $N^*(1/2+)$ state.
Only a rough plateau from time slice 7 to 9 can be found and the mass of 
$N^*(3/2+)$ is extracted from this time window so the results for this state 
should be taken with caution. To access later 
time slices, a large number of configurations are needed to increase the 
signal-to-noise ratio. 
The signal for $N^*(3/2-)$, on the other hand, is much stronger since 
it is the dominant component in the negative parity channel. 
A nice flat area can be found between 
time slices 8 and 12. The mass of the $N^*(3/2-)$ is extracted from 
time slice 10 to 12. 

\begin{figure}
\centerline{\psfig{file=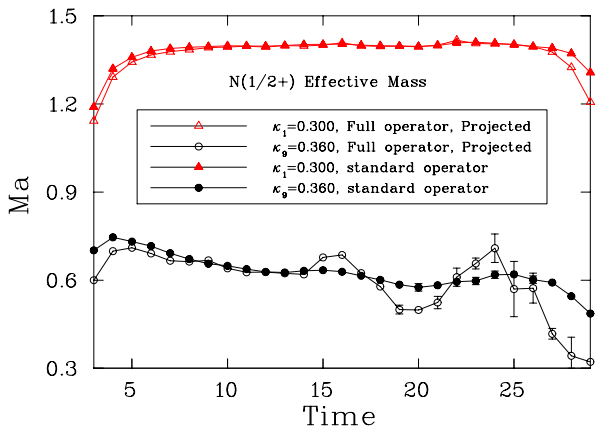,width=0.90\hsize,angle=90}}
\caption{Comparison of nucleon effective masses at the smallest and largest pion masses 
from two ways: one from the standard operator in Eq.(~\protect\ref{nuc1}), 
the other from the $1/2+$ component of the full operator in Eq.(~\protect\ref{eq:chifull}).}
\label{lat_eMass_nuc12pos_proj.vs.standard}
\end{figure}

Fig.~\ref{eMass_nuc12} displays the effective masses in the spin-1/2 sector. 
Here $N(1/2+)$ is the dominant component, while $N(1/2-)$ is the weaker one, in 
their respective parity channels.
The $N(1/2+)$ state is extracted from time slice 11 - 14.
For $N(1/2-)$, a rough flat area can be found from time 
slice 7 to 12, but we use 9 to 11 since the value from later time slice has 
smaller systematic error. To show the quality of the nucleon from the two different measurements,
a comparison of the effective masses of the nucleon at two kappa values 
is given in Fig.~\ref{lat_eMass_nuc12pos_proj.vs.standard}.
The projected nucleon at the smallest pion mass shows some instability beyond time slice 15.

\begin{figure}
\centerline{\psfig{file=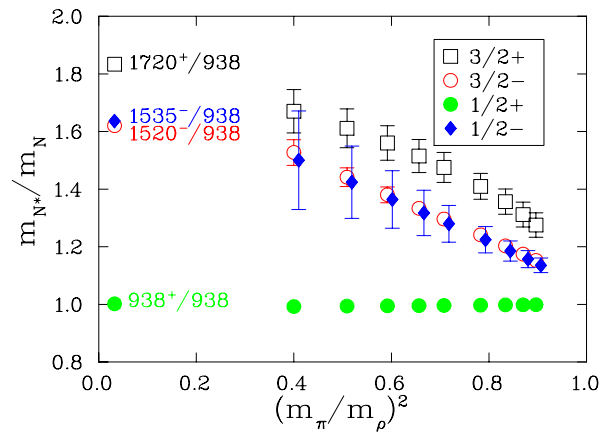,width=0.90\hsize,angle=90}}
\caption{Mass ratio of the projected $N^*$ states to the ground-state nucleon 
as a function of the mass ratio squared $(\pi/\rho)^2$. 
The four lattice states with separated spin-parity are 
symbol-coded (and color-coded if color is visible) with the experimental candidates 
which are indicated on the left at the physical point.}
\label{Ratio3_nuc_all}
\end{figure}

\begin{figure}
\centerline{\psfig{file=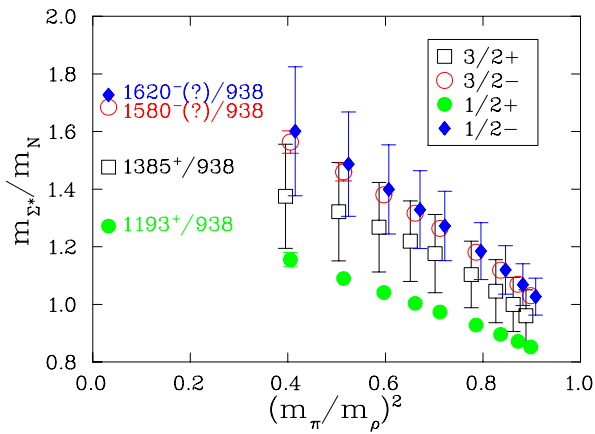,width=0.90\hsize,angle=90}}
\caption{Same as in Fig.~\protect\ref{Ratio3_nuc_all}, but for the $\Sigma$ states.}
\label{Ratio3_sig_all}
\end{figure}

Fig.~\ref{Ratio3_nuc_all} presents results of the mass
ratios extracted from the correlation functions for the $N^*$ 
states to the nucleon ground state as a function of the mass 
ratio $(\pi/\rho)^2$.
We use mass ratios because they have minimal dependence on the uncertainties in
determining the scale and the quark masses, and have smaller 
statistical errors.
In this figure, we slightly shift the points for the $N^*(1/2-)$ state 
to the right hand side to avoid overlapping. 
The mass of the nucleon ground state we use in the mass ratios is 
produced on the same configurations from the standard 
interpolating field. It is encouraging to see that the $1/2+$ nucleon 
obtained from the standard operator agrees with that from the projected 
$1/2+$ nucleon in the spin-3/2 operator (as indicated by the ratio of 1).
Clear splitting is seen between the $N(1/2+)$ state 
and its parity partner $N^*(1/2-)$ state in this figure. 
The two negative-parity states $N^*(1/2-)$ and $N^*(3/2-)$ are 
degenerate within errors, which is consistent with the fact that the corresponding states 
$N(1535)1/2-)$ and $N(1520)3/2-$ in the observed spectrum are very close to each other.
We do similar calculations in the $\Sigma^*$ and $\Xi^*$ channels. 
The results for the $\Sigma^*$ channel is shown in Fig.~\ref{Ratio3_sig_all}.
In the PDG~\cite{pdg04} the two positive-parity states 
$\Sigma(1193)(1/2+)$ and $\Sigma(1385)(3/2+)$ are well established, which 
we identify as $\Sigma(1/2+)$ and $\Sigma^*(3/2+)$ on the lattice, respectively.
The situation in the negative-parity is unclear. The two possible candidates for 
the lattice states $\Sigma^*(1/2-)$ and $\Sigma^*(3/2-)$ are 
2-star states $\Sigma(1620)1/2-$ and $\Sigma(1580)3/2-$, respectively.
In the figure, they have question marks, and the four lattice states are matched one to one 
with the experimental candidates.

\begin{figure}
\centerline{\psfig{file=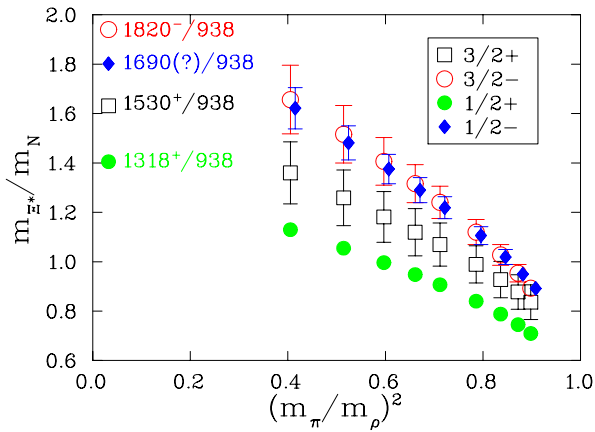,width=0.90\hsize,angle=90}}
\caption{Same as in Fig.~\protect\ref{Ratio3_nuc_all}, but for the $\Xi$ states.}
\label{Ratio3_xi_all}
\end{figure}

A similar situation exists in the $\Xi$ channel shown in Fig.~\ref{Ratio3_xi_all}.
The two well-established positive-parity states $\Xi(1318)1/2+$ and $\Xi(1530)3/2+$ 
are identified with their lattice counterparts.
The two possible candidates for
the negative-parity lattice states $\Xi^*(1/2-)$ and $\Xi^*(3/2-)$ are
3-star $\Xi(1690)$ state with unknown spin-parity and 
the 3-star $\Xi(1820)3/2-$ state, respectively.
All of the mass ratios in this sector are listed in Table \ref{tab:nstar}. 
\begin{table*}
\label{tab:nstar}
\caption{Mass ratios for the four states in the isospin-1/2 and spin-3/2 family 
after parity and spin projections.
The results are from the full  interpolating field in Eq.~(\protect\ref{eq:chifull}).
The numbers in brackets are statistical errors in the last digits.
The last column indicates the time window from which the results are extracted. 
The absolute values for the pion mass and the nucleon mass are also given for 
conversion purposes.}
\begin{tabular}{l|llllllllll}
 \hline
$m_\pi$ (GeV)         &2.10(1)    &1.87(1)    &1.65(1)     & 1.43(1)    & 1.20(1)    & 1.08(1)    & 0.96(1) & 0.82(1) & 0.68(1) & 10-15\\
$m_N$ (GeV)           &3.45(1)    &3.12(1)    &2.80(1)     &2.49(1)     &2.19(1)     &2.04(1)     &1.88(1)  &1.72(2)  &1.56(2)  & 10-15\\
$m_\pi$/$m_\rho$      &0.948(1)   &0.935(1)   &0.916(2)    &0.889(2)    &0.847(2)    &0.817(3)    &0.777(4) &0.723(4) &0.643(5) & 10-15\\
 \hline
$N$(1/2+)/$N$         &1.00(0) &1.00(0) &1.00(0) &1.00(0)  &1.00(0)  &1.00(0)  &1.00(0)  &1.00(1)  &1.00(2) & 11-14\\
$N^*$(1/2-)/$N$       &1.14(3) &1.16(3) &1.19(4) &1.22(5)  &1.28(6)  &1.32(8)  &1.36(10) &1.42(13) &1.50(17)& 9-11\\
$N^*$(3/2+)/$N$       &1.28(4) &1.31(4) &1.36(4) &1.41(4)  &1.48(5)  &1.52(6)  &1.56(6)  &1.61(7)  &1.67(8) & 7-9\\
$N^*$(3/2-)/$N$       &1.15(1) &1.17(1) &1.20(1) &1.24(2)  &1.30(2)  &1.33(2)  &1.38(3)  &1.44(3)  &1.53(4) & 10-12\\
 \hline
$\Sigma^*$(1/2+)/$N$  &0.85(0) &0.87(0) &0.90(0) &0.93(0)  &0.97(0)  &1.00(0)  &1.04(1)  &1.09(2)  &1.16(3) & 15-17\\
$\Sigma^*$(1/2-)/$N$  &1.03(6) &1.07(7) &1.12(8) &1.18(10) &1.27(12) &1.33(14) &1.40(15) &1.49(18) &1.60(22)& 10-12\\
$\Sigma^*$(3/2+)/$N$  &0.96(9) &1.00(9) &1.05(11) &1.10(12) &1.18(14) &1.22(14) &1.27(16) &1.32(17) &1.38(18)& 10-12\\
$\Sigma^*$(3/2-)/$N$  &1.03(1) &1.07(1) &1.12(1) &1.18(2)  &1.26(2)  &1.32(2)  &1.38(3)  &1.46(3)  &1.56(4) & 10-12\\
 \hline
$\Xi^*$(1/2+)/$N$     &0.71(0) &0.75(0) &0.79(1) &0.84(1)  &0.91(1)  &0.95(1)  &1.00(1)  &1.05(1)  &1.13(2) & 11-14\\
$\Xi^*$(1/2-)/$N$     &0.89(2) &0.95(3) &1.02(3) &1.11(4)  &1.22(4)  &1.29(5)  &1.38(6)  &1.48(7)  &1.62(8) & 10-12\\
$\Xi^*$(3/2+)/$N$     &0.84(7) &0.88(7) &0.93(7) &0.99(8)  &1.07(9)  &1.12(10) &1.18(10) &1.26(11) &1.36(13)& 11-13\\
$\Xi^*$(3/2-)/$N$     &0.89(3) &0.95(4) &1.03(4) &1.12(5)  &1.24(7)  &1.32(8)  &1.41(10) &1.52(12) &1.66(14)& 13-15 \\
 \hline
\end{tabular}
\end{table*}


\section{The Issue of Interpolating Fields}
\label{sec:diff}
The computational cost of evaluating each Lorentz combination
in Eq.~(\ref{eq:chifull}) is relatively high, about 100 times that for the
ground-state nucleon. Therefore, it took a long time
for us to get the 100 configurations. 
In Ref.~\cite{Zanotti:2003fx}, only the first term of the 
full interpolating field,
\begin{equation}
\chi_\mu = \epsilon_{abc}\left( {u^a}^{T}C\gamma_5\gamma_\mu d^b\right)
 \gamma_5 u^c,
\label{eq:firsttermonly}
\end{equation}
was considered.
In our calculation, we deliberately separated the contributions from 
individual terms, so we are in a position to 
investigate possible differences between the two interpolating fields.


 \begin{figure}
 \centerline{\psfig{file=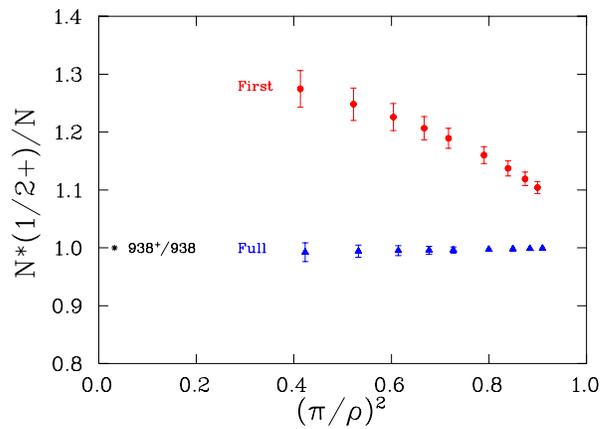,width=0.90\hsize,angle=90}}
 \caption{The mass difference between the result from the full
 interpolating field and that from the first term of
 the interpolating field for the $N(1/2+)$ state.}
 \label{fig:ratio12_pos}
 \end{figure}

We found that the projected spin-3/2 components are the same within 
statistical errors.
In the spin-1/2 sector, however, the situation is different.
Fig.~\ref{fig:ratio12_pos} shows the comparison in the $1/2+$ channel.
There is a significant difference (up to 30\%) between these 
two results. Similar difference is found in the $1/2-$ channel.
This result is different from that in Ref.~\cite{Zanotti:2003fx}.
So as far as spin-3/2 states are concerned, the first 
term of the interpolating field can be used to do the calculation. 
However, if the masses of spin-1/2 states are desired from such interpolating 
fields via spin projection, 
the full interpolating field is required.  


\section{Isospin-3/2 and Spin-3/2 Baryons}
\label{sec:spin32}

For a $\Delta^+$ state, the interpolating field is
\begin{equation}
\label{eq:deltainter}
\chi_\mu^{\Delta^+} =\frac{1}{\sqrt{3}} \epsilon_{abc}
[2 (u^{Ta} C \gamma^\mu d^b ) u^c + (u^{Ta} C \gamma^\mu u^b) d^c ].
\end{equation}
Interpolating fields for other decuplet baryons can similarly be 
obtained by appropriate substitutions of quark fields.

First of all, we would 
like to show the correlation functions since these are the bases 
of mass extraction. 
\begin{figure}
\parbox{.5\textwidth}{%
\centerline{\psfig{file=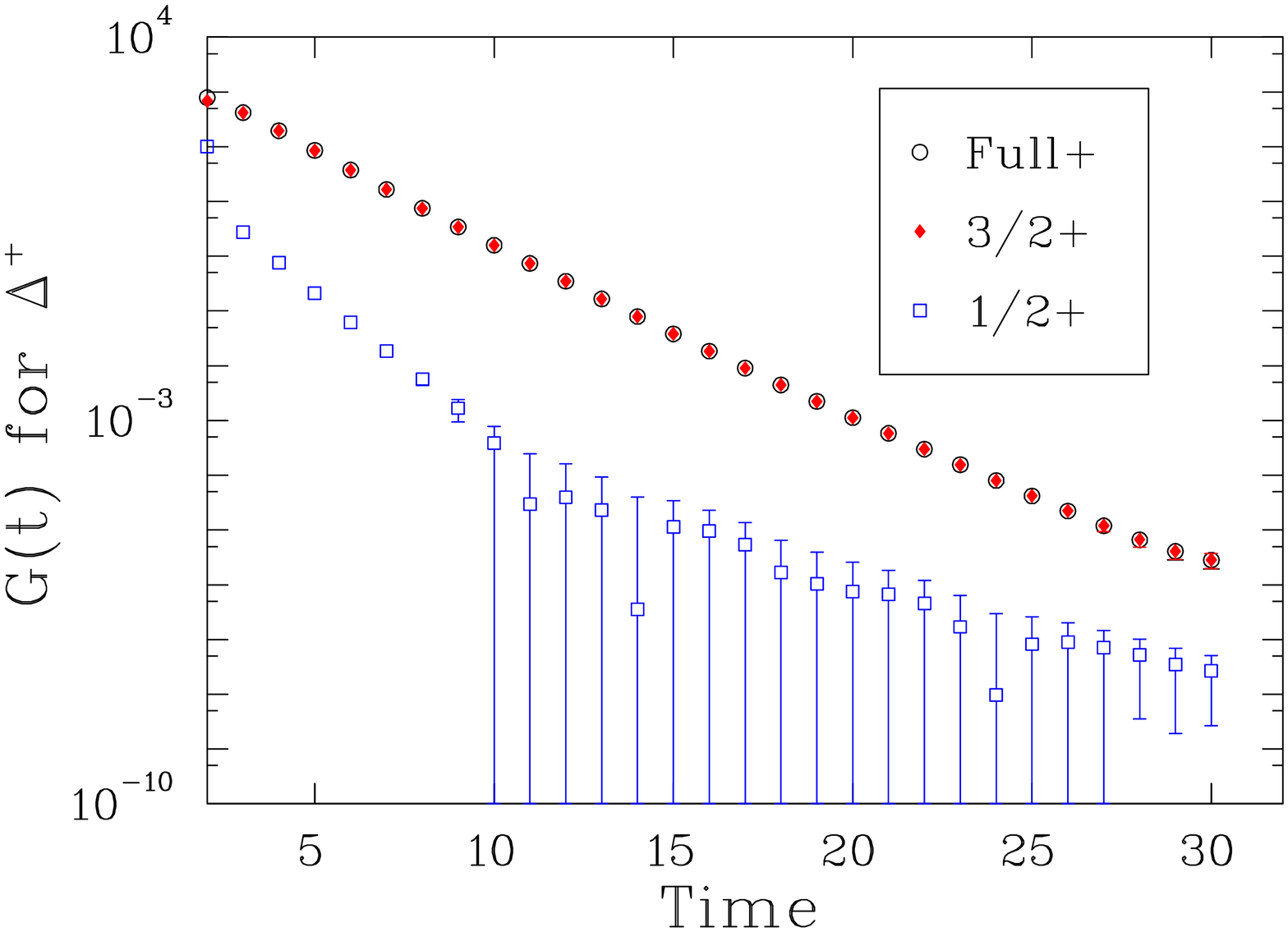,width=0.90\hsize,angle=90}}
}\hfill
\parbox{.5\textwidth}{%
\centerline{\psfig{file=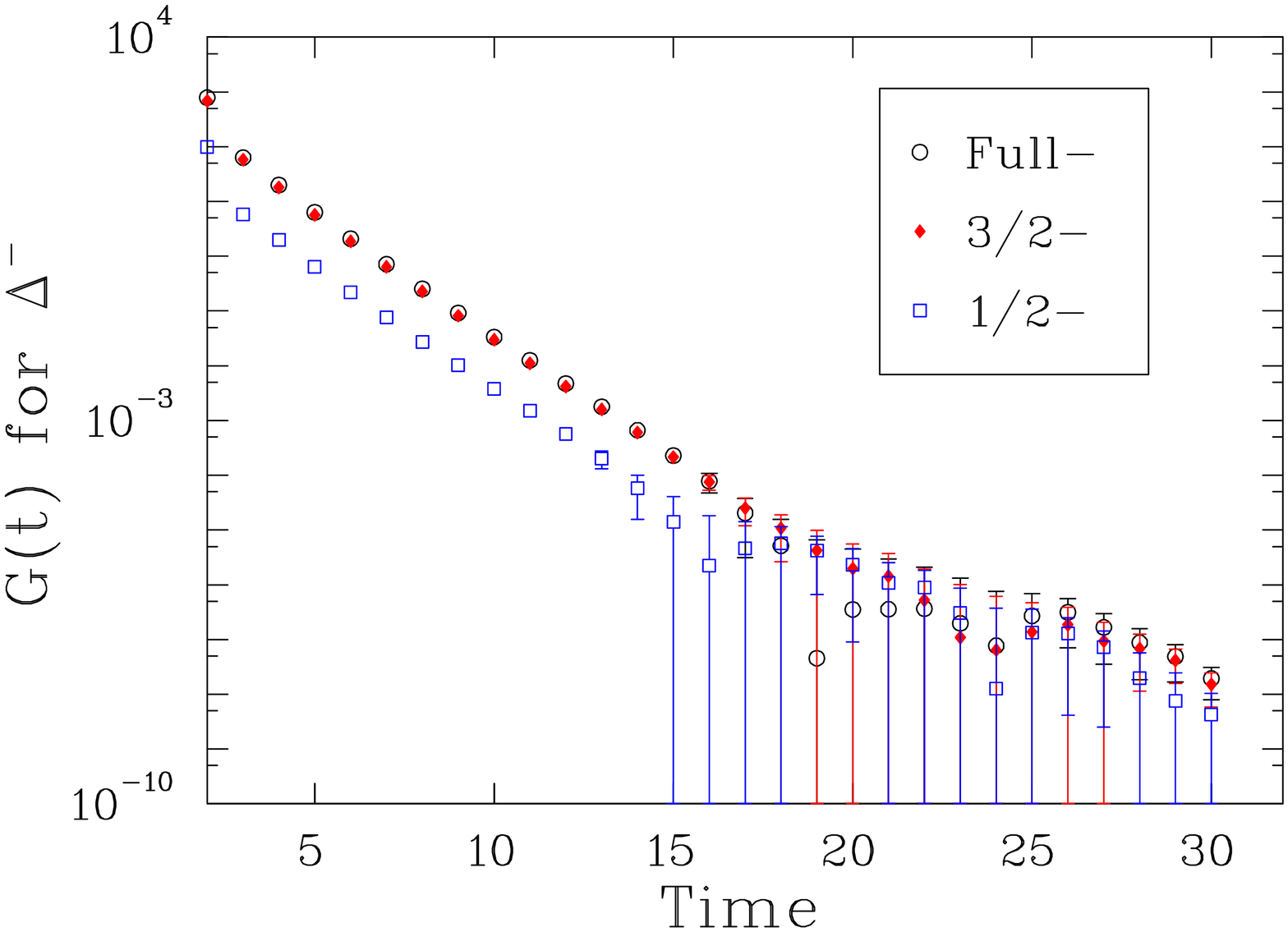,width=0.90\hsize,angle=90}}
}
\caption{The various correlation functions 
(unprojected, spin-3/2 projected, spin-1/2 projected) 
for the Delta states in the positive-parity (top) and negative-parity (bottom) channels 
at the smallest quark mass.}
\label{corr_del32}
\end{figure}
Fig.~\ref{corr_del32} presents the correlation
functions for $\Delta$ states in both parity channels.
Again, the relation in Eq.~(\ref{sum}) is satisfied. 
The spin-3/2 component almost completely dominates in both channels. 
The $1/2+$ component is weaker by several orders of magnitude.
This indicates 
that the interpolating field in Eq. \ref{eq:deltainter} has a small overlap 
with spin-1/2 states. 
Nonetheless, we could extract a discernable signal with 300 configurations.
The mass of the spin-3/2 
states can be extracted from relatively large time slices,
while the mass of the spin-1/2 states are extracted 
before time slice 10 since the data becomes 
very noisy beyond that point.
Another observation is that 
$\Delta(1/2+)$ is heavier than $\Delta(3/2+)$, while
$\Delta(1/2-)$ is about the same as $\Delta(3/2-)$.
  
\begin{figure}
\centerline{\psfig{file=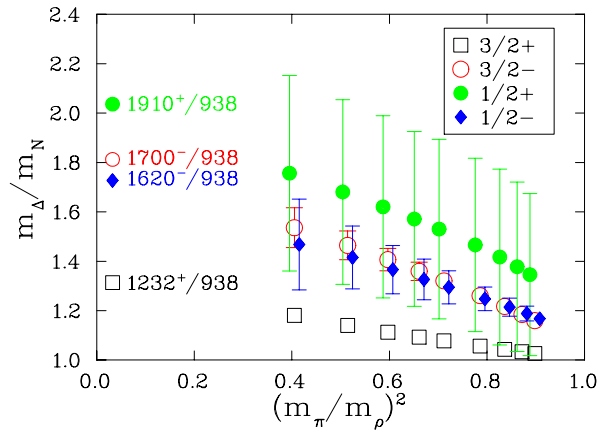,width=0.90\hsize,angle=90}}
\caption{Mass ratio of the projected $\Delta$ states to the ground-state nucleon 
as a function of the mass ratio squared $(\pi/\rho)^2$. 
The four lattice states with separated spin-parity are 
symbol-coded with the experimental candidates 
which are indicated on the left at the physical point.}
\label{fig:delta32_all}
\end{figure}

As usual, the fitting is done on the effective masses,
using the same spin and parity projection techniques.
Here we show the results directly without showing the effective mass plots.
Fig.~\ref{fig:delta32_all} shows the results in the  $\Delta$  channel.
The trend of the $\Delta(3/2+)/N$ data points with decreasing quark masses
is clearly toward the observed ratio of $\Delta(1232)/N(938)$.
The splitting between $\Delta(3/2-)$ and $\Delta(3/2+)$ is consistent with 
that in experiment (470 MeV).
There is a small hint that $\Delta(3/2-)$ lies above $\Delta(1/2-)$, but 
the errors are too big to resolve the two states 
which are close to each other in experiment.
For better viewing, the points for the 
$\Delta(1/2-)$ state have been slightly shifted to the right hand side, and the points
for the $\Delta(1/2+)$ state have been shifted to the left hand side. The large errors show 
that this state is difficult to extract with limited statistics.
The splitting between $\Delta(1/2+)$  and its parity partner $\Delta(1/2-)$ also appears 
consistent with that in experiment (290 MeV). The signal 
of this $\Delta(1/2+)$ state is the weakest in the four $\Delta$ states. 
In all, the splitting pattern of these states is consistent with that observed in experiment
and with .Ref.~\cite{Zanotti:2003fx},
despite weak signals for the $\Delta(1/2\pm)$ states.

\begin{figure}
\centerline{\psfig{file=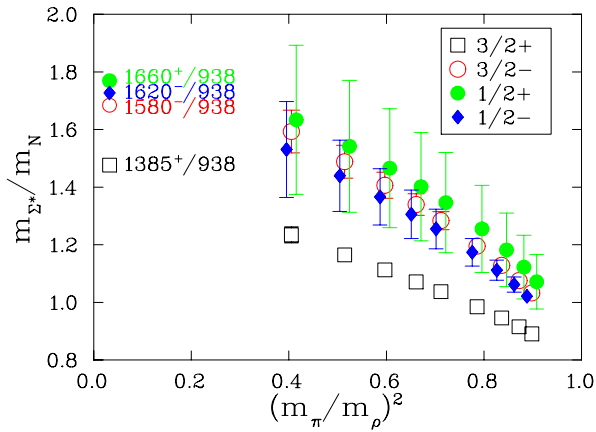,width=0.90\hsize,angle=90}}
\caption{Same as in Fig.~\protect\ref{fig:delta32_all}, but for the $\Sigma$ states.}
\label{fig:dsigma32_all}
\end{figure}

\begin{figure}
\centerline{\psfig{file=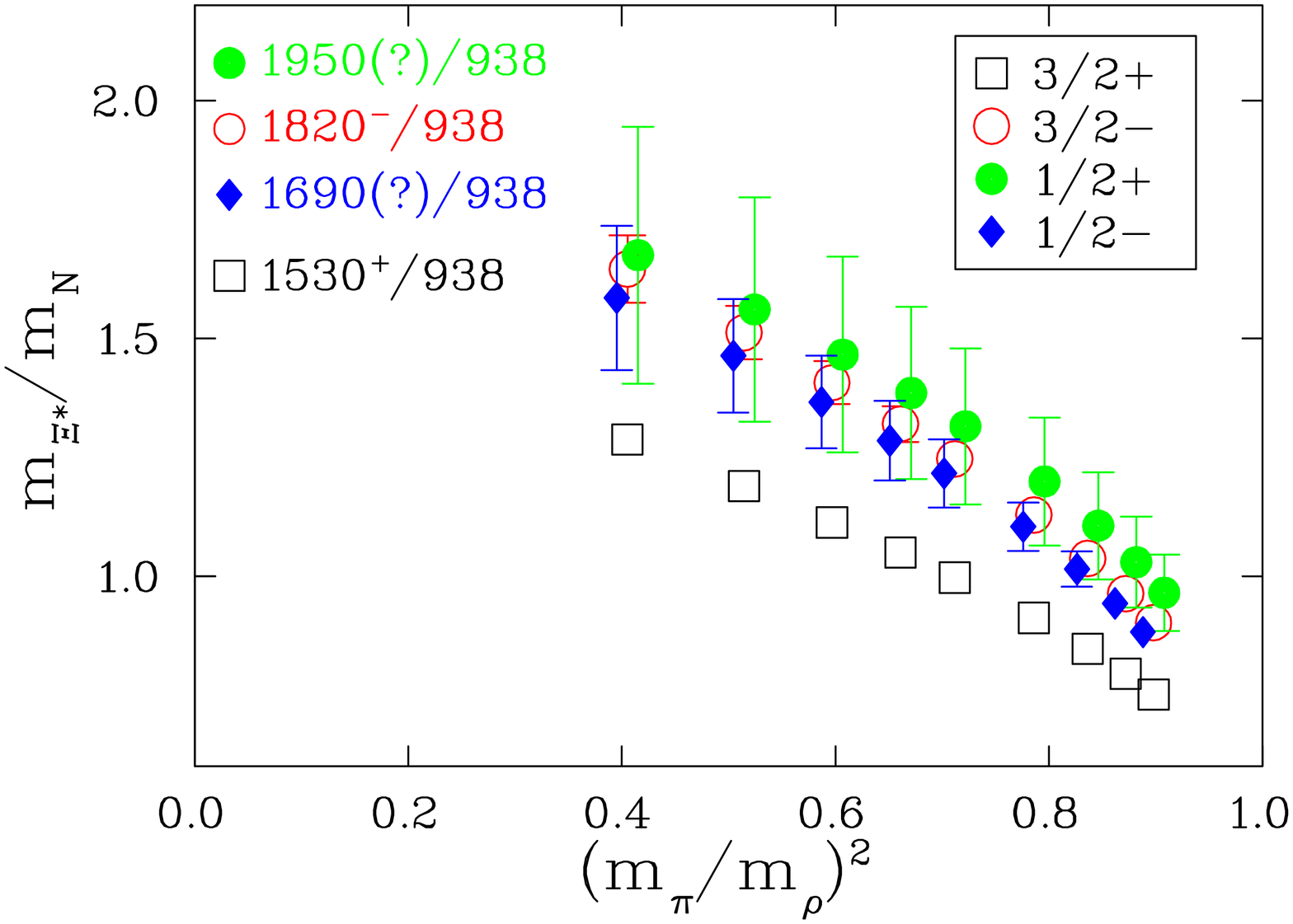,width=0.90\hsize,angle=90}}
\caption{Same as in Fig.~\protect\ref{fig:delta32_all}, but for the $\Xi$ states.}
\label{fig:dxi32_all}
\end{figure}

Similar patterns exhibit in the $\Sigma^*$
and $\Xi^*$ states in the decuplet, as shown in 
Fig.~\ref{fig:dsigma32_all} and Fig.~\ref{fig:dxi32_all}.
Strong signals and stable results for the spin-3/2 states, 
relatively weak signals and big error bars for the spin-1/2 states.
Note that there is a systematic improvement in the signal for the $1/2+$ states as 
the number of the strange quarks increases from 0 in $\Delta$ to 1 in $\Sigma^*$ to 2 in $\Xi^*$.
This can be attributed to the stabilizing effects of the heavier strange quark.
The experimental situation for the $\Xi^*$ states is not clearly settled yet~\cite{pdg04}.
That is why we put question marks on some of the states.
The $\Xi(1820)(3/2-)$ state has 3-star status which we identify with our $3/2-$ state on the lattice.
There is a 3-star state $\Xi(1690)$ with unknown spin-parity. We identify this state with our 
$1/2-$ state on the lattice.
Our highest state $\Xi^*(1/2+)$ is identified with the 3-star state of $\Xi(1950)$ which 
also has unknown spin-parity in the Particle Data Group.
All of our results in this sector are summarized in Table~\ref{tab:dec32}. 

\begin{table*}
\label{tab:dec32}
\caption{Mass ratios for the four states in the isospin-3/2 and spin-3/2 family 
(baryon decuplet) after parity and spin projections.
The numbers in brackets are statistical errors in the last digits.
The last column indicates the time window from which the results are extracted.}

\begin{tabular}{l|lllllllllll}
 \hline
$m_\pi$/$m_\rho$      &0.948(1)   &0.935(1)   &0.916(2)    &0.889(2)    &0.847(2)    &0.817(3)    &0.777(4) &0.723(4) &0.643(5) & 10-15\\
 \hline
$\Delta$(1/2+)/$N$    &1.35(33) &1.38(34) &1.42(36) &1.47(35) &1.53(36) &1.57(35) &1.62(37) &1.68(37) &1.76(40) &7-9 \\
$\Delta$(1/2-)/$N$    &1.17(2)  &1.19(3)  &1.21(4)  &1.25(5)  &1.29(7)  &1.33(8)  &1.37(10) &1.42(13) &1.47(18) &10-12\\
$\Delta$(3/2+)/$N$    &1.03(1)  &1.03(1)  &1.04(1)  &1.06(1)  &1.08(1)  &1.09(2)  &1.11(2)  &1.14(2)  &1.18(3)  &11-13\\
$\Delta$(3/2-)/$N$    &1.16(1)  &1.18(2)  &1.22(2)  &1.26(2)  &1.32(3)  &1.36(4)  &1.41(5)  &1.46(6)  &1.54(8)  &11-13\\
 \hline
$\Sigma^*$(1/2+)/$N$  &1.07(9)  &1.12(11) &1.18(13) &1.26(15) &1.35(17) &1.40(19) &1.47(21) &1.54(23) &1.63(26) &7-9 \\
$\Sigma^*$(1/2-)/$N$  &1.02(2)  &1.06(3)  &1.11(3)  &1.17(5)  &1.26(7)  &1.31(8)  &1.37(10) &1.44(12) &1.53(17) &10-12 \\
$\Sigma^*$(3/2+)/$N$  &0.89(1)  &0.92(1)  &0.95(1)  &0.98(1)  &1.04(1)  &1.07(2)  &1.11(2)  &1.16(2)  &1.23(3)  &11-13 \\
$\Sigma^*$(3/2-)/$N$  &1.03(1)  &1.08(2)  &1.13(2)  &1.20(2)  &1.28(3)  &1.34(4)  &1.41(5)  &1.49(6)  &1.59(7)  &11-13 \\
 \hline
$\Xi^*$(1/2+)/$N$     &0.96(8)  &1.03(10) &1.11(11) &1.20(13) &1.32(16) &1.39(18) &1.47(21) &1.56(24) &1.68(27) &7-9   \\
$\Xi^*$(1/2-)/$N$     &0.88(2)  &0.94(3)  &1.02(4)  &1.10(5)  &1.22(7)  &1.29(8)  &1.37(10) &1.46(12) &1.58(15) &10-12 \\
$\Xi^*$(3/2+)/$N$     &0.75(1)  &0.79(1)  &0.85(1)  &0.91(1)  &1.00(1)  &1.05(2)  &1.11(2)  &1.19(2)  &1.29(3)  &11-13 \\
$\Xi^*$(3/2-)/$N$     &0.90(1)  &0.96(2)  &1.04(2)  &1.13(2)  &1.25(3)  &1.32(4)  &1.41(5)  &1.51(6)  &1.65(7)  &11-13  \\
 \hline
\end{tabular}
\end{table*}

\section{Summary and Outlook}
\label{sec:summary}

In this exploratory study, 
we have computed the mass spectrum of spin-3/2 baryons
using the method of quenched QCD on an anisotropic lattice.
The full isospin-1/2 and spin-3/2 interpolating field in Eq.~(\ref{eq:chifull})
is used.
We analyzed 100 configurations despite a big increase 
in computing demand as compared to a truncated version of the interpolating field. 
Four states with definite spin-parity are isolated for each particle type 
using parity projection and spin projection.
The need for spin projection is clearly demonstrated in the positive-parity 
channel whose correlation function is dominated by the spin-1/2 component.
Clear signals are obtained for both the spin-projected $N^*(3/2\pm)$
and the $N^*(1/2\pm)$ states, although the latter are usually weaker, 
resulting in relatively large errors. 
The results in the $\Sigma^*$ and $\Xi^*$ channels are reported for the first time. 
Some of our lattice result can be considered as predictions in cases where 
the spin-parity assignment is unknown in the PDG.
The spin-1/2$\pm$ states extracted from the 
spin-3/2 interpolating fields are in good agreement with 
those from the standard spin-1/2 interpolating fields, providing a non-trivial 
check of the calculation.
Furthermore, we find that the projected spin-1/2 states are quite different 
in the full and the truncated interpolating fields.
This means that as far as spin-3/2 states are concerned, the first 
term of the interpolating field is sufficient. 
However, if the masses of spin-1/2 states are computed from such interpolating 
fields via spin projection, the full interpolating field is required.  

As an independent check of the calculation,
we carried out a parallel calculation of the usual baryon decuplet 
(isospin-3/2 and spin-3/2) on the same lattice with 300 configurations, 
using the same projection techniques.
The pattern in the $\Delta$ states is consistent 
with the previous calculation~\cite{Zanotti:2003fx} and with experiment.
This reinforces the efficacy of the methods used in separating the spins and parities.
The results in the $\Sigma^*$ and $\Xi^*$ channels are new and are used 
to shed light on the spin-parity of some states in the PDG.

Having established the signals and the methods used to isolate the $3/2\pm$ and 
$1/2\pm$ states,
improvement can be made in a number of areas in future studies. 
First, higher statistics (probably on the order of 1000 configurations) are needed 
to beat down the errors in the weaker spin-projected states. 
Second, smaller pion masses are desired to perform a chiral extrapolation and 
make better contact with experiment.
Third, both the lattice spacing and the box size should be 
varied to assess possible discretization effects. 
In the long run, the calculations should be done with dynamical gauge configurations 
in order to assess the effects of quenching in this sector.

\begin{acknowledgments}
This work is supported in part by U.S. Department of Energy under grant DE-FG02-95ER40907,
and computing resources at NERSC.
FXL would like to thank the Institute for Nuclear Theory at the University of Washington for
its hospitality during the completion of this work.
\end{acknowledgments}


\end{document}